\documentstyle[preprint,aps]{revtex}

\begin{document}
\draft
\title {Direct neutron capture of $^{48}$Ca at $kT = 52~{\rm{keV}}$}
\author{P.~Mohr and H.~Oberhummer}
\address{Institut f\"ur Kernphysik, Wiedner Hauptstr.~8-10, TU Wien,
	A-1040 Vienna, Austria}
\author{H.~Beer}
\address{Forschungszentrum Karlsruhe, Institut f\"ur Kernphysik III,
	P.~O.~Box 3640, D-76021 Karlsruhe, Germany}
\author{W.~Rochow, V.~K\"olle, and G.~Staudt}
\address{Physikalisches Institut, Universit\"at T\"ubingen,
	Auf der Morgenstelle 14, D-72076 T\"ubingen, Germany}
\author{P.~V.~Sedyshev and Yu.~P.~Popov}
\address{Frank Laboratory of Neutron Physics, JINR,
	141980 Dubna, Moscow Region, Russia}
\date{\today}
\maketitle
\begin{abstract}
The neutron capture cross section of $^{48}$Ca was measured 
relative to the known gold cross section
at $kT = 52~{\rm{keV}}$ using the fast cyclic activation technique.
The experiment was performed at the Van-de-Graaff accelerator,
Universit{\"a}t T{\"u}bingen. The new experimental result is in
good agreement with a calculation using the direct capture model.
The $1/v$ behaviour of the capture cross section at thermonuclear
energies is confirmed, and the adopted reaction rate which is based
on several previous experimental investigations remains unchanged.
\end{abstract}



\narrowtext

\section{Introduction}\label{s1}
The neutron capture cross section of the neutron--rich and doubly magic
nucleus $^{48}$Ca is dominated by the direct capture (DC) mechanism
at thermonuclear energies \cite{kra96,beer96}. This capture mechanism has to be
well understood for the analysis of neutron--induced nucleosynthesis
in the vicinity of $^{48}$Ca which is also of relevance for the
Ca--Ti isotopic abundance anomalies in certain 
primitive meteorites \cite{san82,zie85,woe93}.

The non--resonant DC cross section was measured by 
Beer {\it et al.} \cite{beer96} and
K\"appeler {\it et al.} \cite{kae85} using activation techniques. 
Weak resonances were found by Carlton {\it et al.} \cite{car87} 
using the time--of--flight (TOF) technique, and the thermal capture
cross section was measured by Beer {\it et al.} \cite{beer96} and
Cranston and White \cite{cra71}. Summarizing the previous experiments,
in the experimentally analyzed energy range
the $^{48}$Ca(n,$\gamma$)$^{49}$Ca cross section shows an almost pure
$1/v$ behaviour which is typical for $s$--wave capture.
However, there is some evidence for a deviation from the $1/v$ law
at $kT = 25~{\rm{keV}}$ in Ref.~\cite{beer96} where the experimental
value is roughly 25\% smaller 
than the calculated cross section
(corresponding to 2.8 times the uncertainty of
the experiment).
This behaviour can be explained 
either by the uncertainty of the experiment or by a destructive
interference between the non--resonant DC and a $1/2^+$ $s$--wave
resonance at $E \approx 1.5~{\rm{keV}}$ \cite{beer96}
which should have been observed in the TOF experiments of Ref.~\cite{car87}.

The most important energies for astrophysical scenarios are in the order
of 10 to 100 keV. Therefore we measured the neutron capture cross
section of $^{48}$Ca using a quasi--Maxwellian neutron energy spectrum
with $kT = 52~{\rm{keV}}$ which is obtained by bombarding a thick
tritium target with protons at $E_p = 1091~{\rm{keV}}$
(72 keV above the T(p,n) threshold at $E_p = 1019~{\rm{keV}}$) \cite{kae87}.

The experimental setup using the fast cyclic activation technique
is close to the one described in Refs.~\cite{beer96,beer94}.
The sample is irradiated for a period $t_{\rm b} = 149.15 \pm 0.01~{\rm{s}}$, 
after this irradiation time
the sample is moved to the counting position in front of a
high--purity germanium (HPGe) 
detector ($t_{\rm{w_1}} = 0.80 \pm 0.01~{\rm{s}}$),
the $\gamma$--rays following the $\beta$--decay of $^{49}$Ca are
detected for a time interval $t_{\rm c} = 149.20 \pm 0.01~{\rm{s}}$, 
and finally the sample
is moved back to the irradiation position 
($t_{\rm{w_2}} = 0.85 \pm 0.01~{\rm{s}}$).
The whole cycle with duration 
$T = t_{\rm b} + t_{\rm{w_1}} + t_{\rm c} + t_{\rm{w_2}} = 300~{\rm{s}}$
is repeated many times to gain statistics.

The sample material consisted of CaCO$_3$ powder
enriched by 77.87~\% in $^{48}$Ca.
Two samples were available with a diameter $d = 6~{\rm{mm}}$
and masses $m_1 = 48.83~{\rm{mg}}$ and $m_2 = 49.06~{\rm{mg}}$.
The samples were sandwiched between two thin gold foils
($m \approx 16~{\rm{mg}}$ per foil), and both the sample
and the two gold foils were put into a cylindrical polyethylen container
with an inner diameter of 6 mm.

The experiment was performed at the 3.5 MeV Van--de-Graaff accelerator
{\sc Rosenau} at the University of T\"ubingen where a tritum target
with an activity of 5 Ci is available for the neutron production
with the $^3$H(p,n)$^3$He reaction. The thickness of the titanium layer
which contains the tritium
is about 150 keV ??? at $E_p = 1091~{\rm{keV}}$.

The neutron flux was limited by the relatively low
beam current of about $15~{\rm \mu A}$. Using this current the temperature
of the tritium target remained below $40^o$ C, and a constant ratio between
the neutron flux and the proton beam current could be observed during 
the whole experiment. Furthermore, the depth profile of the tritium target
was constant during the experiment. This
was controlled by measuring the neutron yield at different energies between
the (p,n) threshold and $E_p = 1091~{\rm{keV}}$.
Changes in the depth profile would affect the neutron energy spectrum. 
The resulting uncertainty of the capture cross section is estimated 
to be smaller than 5\%.

Because of the relatively low neutron flux
several improvements had to be
achieved compared to our previous experiment \cite{beer96}.
The improvements are 
\begin{itemize}
\item a better detection efficiency
using a HPGe detector with 98.6\% relative efficiency
\item a longer distance
(1 m) between irradiation and detection position which is used for a better
shielding of the HPGe detector
(75 cm lithium--loaded paraffine between the neutron source and the
HPGe detector, and 0.3 cm cadmium and 15 cm lead around the HPGe detector) 
\end{itemize}
A typical spectrum is shown in Fig.~\ref{fig:spec}. 
In the insets the
relevant areas around 411.8 keV (decay of $^{198}$Au),
and 3084.4 and 4071.9 keV (decay of $^{49}$Ca) are shown. 

The accumulated number of counts from a total of $N$ cycles,
$C=\sum_{i=1}^n C_i$, where $C_i$,
the counts after the i-th cycle, are calculated for a chosen irradiation
time, $t_{\rm b}$, which is short
enough compared with the fluctuations of the neutron flux, 
is~\cite{beer94}
\begin{equation}
\label{eq1}
C  = \epsilon_{\gamma}K_{\gamma}f_{\gamma}\lambda^{-1}[1-\exp(-\lambda
t_{\rm c})]
\exp(-\lambda t_{\rm{w_1}})
 \frac{1-\exp(-\lambda t_{\rm b})}{1-\exp(-\lambda T)} N \sigma_\gamma
{[1-f_{\rm b} \exp(-\lambda T)]}
 \sum_{i=1}^n \Phi_i
 \end{equation}
with
\begin{displaymath}
f_{\rm b}  =  \frac{\sum_{i=1}^n \Phi_i \exp[-(n-i)\lambda T]}{
\sum_{i=1}^n \Phi_i} \quad .
\end{displaymath}
The following additional quantities have been defined;
$\epsilon_\gamma$: Ge-efficiency, $K_\gamma$:
$\gamma$-ray absorption, $f_\gamma$: $\gamma$-ray intensity per decay,
$N$: the thickness (atoms per barn) of target
nuclei, $\sigma_\gamma$: the capture cross-section, $\Phi_i$: the neutron flux
in the i-th cycle. The
quantity $f_{\rm b}$ is calculated from the 
registered flux history of a $^6$Li
glass monitor.

The efficiency of the HPGe-detector was determined
by a calculation using the computer code GEANT \cite{GEANT}.
The results of the detector simulation 
were tested experimentally by measuring the relative
intensities of $\gamma$--rays following the decays of $^{115,117}$Cd.
For this purpose a sample of $^{\rm nat}$Cd with the same geometry
as the calcium samples (diameter $d = 6~{\rm{mm}}$, 
mass $m = 343.47~{\rm{mg}}$) 
was mounted exactly in the same way as for the calcium experiment.
Additionally, the capture $\gamma$--rays of $^{27}$Al(p,$\gamma$)$^{28}$Si
at the resonance at $E_p = 1317~{\rm{keV}}$
which are well--known in literature \cite{brenn95,endt90} 
were measured using a thick
$^{27}$Al target  and a ``D''-shaped
scattering chamber which was especially designed for the detection of
capture $\gamma$--rays \cite{luessem}.
The $\gamma$--ray absorption coefficients in the sample and in the gold foils
were calculated using the tables provided by National Nuclear Data Center,
Brookhaven National Laboratory, via WWW, and based on Ref.~\cite{cullen}.
These corrections are in the order of a few \% or even less for the
$\gamma$--lines from the $^{49}$Ca decay.
The half--lives and the $\gamma$--ray intensities per decay of $^{49}$Ca
and $^{198}$Au are given in Table~\ref{tab:tab1}.

Using Eq.~\ref{eq1} the neutron capture cross section of $^{48}$Ca
can be determined relative to the known activation cross section
of $^{197}$Au which was calculated at $kT = 52~{\rm{keV}}$ in 
Ref.~\cite{kae87} based on experimental data of Ref.~\cite{mack75}:
$\sigma (^{197}{\rm{Au}}) = 431 \pm 15~{\rm{mb}}$. The experimental
result for $^{48}$Ca at $kT = 52~{\rm{keV}}$ is
$\sigma_{\rm{exp}}(^{48}{\rm{Ca}}) = 623 \pm 50~{\rm{\mu b}}$.
The following uncertainties were taken into account:
statistical ($1.5\%$),
neutron energy variations due to tritium density changes 
in the target ($< 5\%$),
enrichment of $^{48}$Ca in the CaCO$_3$ sample ($2.4\%$),
$\gamma$ branching ratios ($1\%$ for the 3084.4 keV line, 
$10\%$ for the 4071.9 keV line),
calculated efficiency of the HPGe detector ($3\%$),
monitoring of neutron flux using the $^6$Li glass monitor ($1\%$),
corrections of the divergence of the neutron beam ($2\%$),
cross section of the gold reference ($3.4\%$),
unknown uncertainties ($2\%$).
Minor uncertainties from the mass determination of the CaCO$_3$ sample
and the gold reference ($<0.1\%$), from the halflives of $^{49}$Ca and
$^{198}$Au ($< 0.1\%$), and from the measuring and waiting times
$T = t_{\rm b} + t_{\rm{w_1}} + t_{\rm c} + t_{\rm{w_2}} = 300~{\rm{s}}$ 
($< 0.1\%$)
can be neglected. The total uncertainty of about $8\%$
is given by the quadratic sum of the above uncertainties.

The experimental value is is good agreement with our theoretical prediction
based on a DC calculation: 
$\sigma_{\rm{calc}}(^{48}{\rm{Ca}}) = 655~{\rm{\mu b}}$ \cite{beer96}.
In Fig.~\ref{fig:sigma} the new experimental result is shown together
with previous experiments \cite{beer96,kae85}
and our DC calculation \cite{beer96}.

In conclusion, the $1/v$ behaviour of the neutron capture cross section
of $^{48}$Ca at thermonuclear energies was confirmed. At $kT = 52~{\rm{keV}}$
the measured capture cross section agrees very well with the calculated value.
The energy-independent reaction rate factor
$N_A < \sigma v > = 1.19 \times 10^5~{\rm{cm^3\, mole^{-1}\, s^{-1}}}$
given in Ref.~\cite{beer96} remains unchanged.

\acknowledgements
We thank the technicians M.~Brandt and G.~Rupp for their help during the
preparation of the experiment, and we thank the Institut f\"ur Strahlenphysik,
Universit\"at Stuttgart, for the borrowing of the HPGe detector. 
This work was supported by Fonds zur F\"orderung der
wissenschaftlichen Forschung in \"Osterreich (project S7307--AST),
Deutsche Forschungsgemeinschaft (DFG) (project Mo739),
and Volkswagen--Stiftung (Az: I/72286).

\begin{table}
\caption{\label{tab:tab1} 
Sample characteristics and decay properties of the
product nuclei $^{49}$Ca and $^{198}$Au.}
\begin{center}
\begin{tabular}{ccccccc}
Isotope		& Chemical	&Isotopic
	& Residual 	& $T_{1/2}$ 		& $E_\gamma$ 
	& Intensity per decay \\
  	& form   		&  composition (\%)
	& nucleus  	& (min)   		& (keV)    
	& (\%) \\
\hline
$^{48}$Ca	& CaCO$_3$	& $77.87 \pm 1.90$	
	&$^{49}$Ca	&8.716$\pm$0.011	& 3084.4 
	&92.1$\pm$1.0		\\
		&		&
	&		&			& 4071.9
	& 7.0$\pm$0.7			\\
$^{197}$Au	&metallic	& 100
	&$^{198}$Au	&2.69\,d		& 411.8
	& 95.50$\pm$0.096	\\
\end{tabular}
\end{center}
\end{table}

\begin{figure}
\caption{
	\label{fig:spec} 
	Energy spectrum of the HPGe detector measured during
	the activation of $^{48}$Ca.
	In the insets the
	relevant areas around 411.8 keV (decay of $^{198}$Au),
	and 3084.4 and 4071.9 keV (decay of $^{49}$Ca) are shown. 
}
\end{figure}

\begin{figure}
\caption{
	\label{fig:sigma} 
	Experimental neutron capture cross section of $^{48}$Ca
	($\bullet$, this work; 
	$\Box$, Ref.~\protect\cite{beer96};
	$\triangle$, Ref.~\protect\cite{kae85})
	compared to a DC calculation 
	(full line, Ref.~\protect\cite{beer96}).
}
\end{figure}

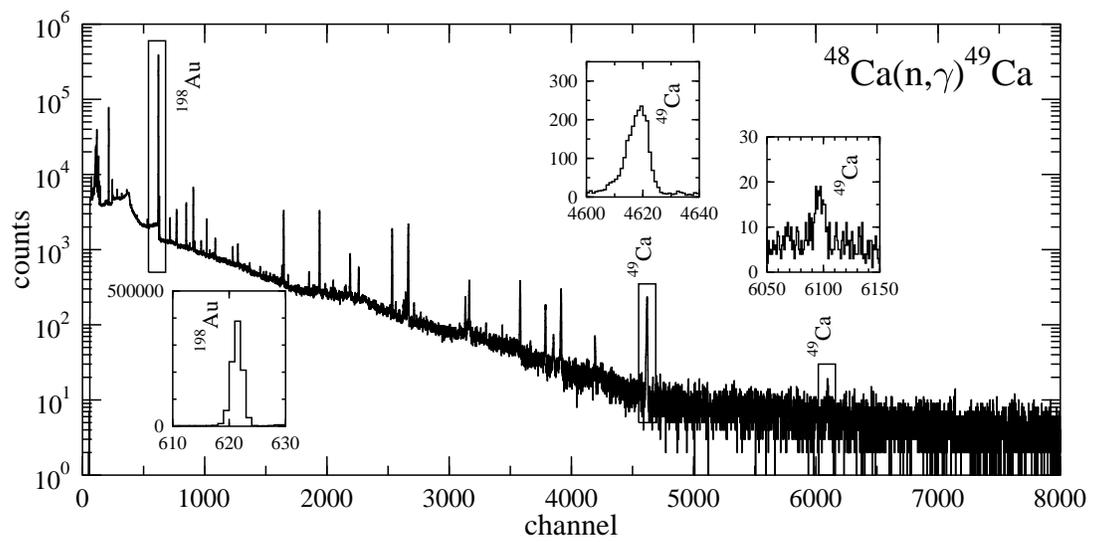
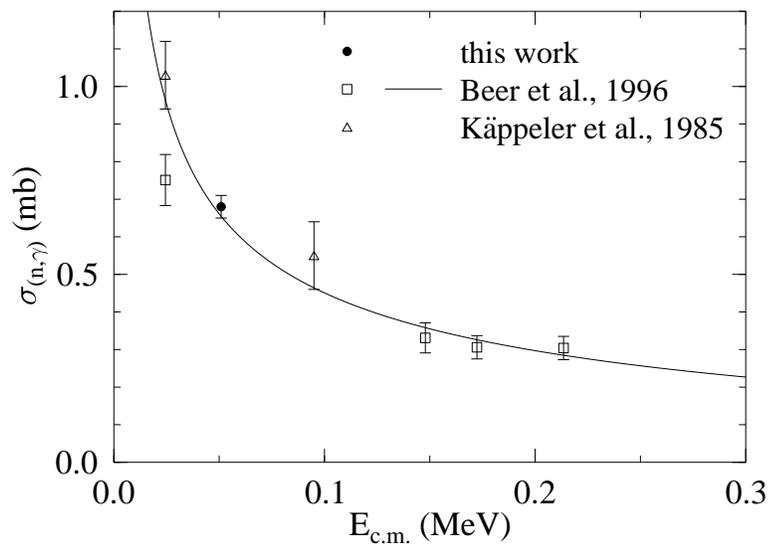
\end{document}